# Impact of Different Desired Velocity Profiles and Controller Gains on Convoy Driveability of Cooperative Adaptive Cruise Control Operated Platoons

Santhosh Tamilarasan, Levent Guvenc
Automated Driving Lab, Ohio State University

## Abstract

As the development of autonomous vehicles rapidly advances, the use of convoying/platooning becomes a more widely explored technology option for saving fuel and increasing the efficiency of traffic. In cooperative adaptive cruise control (CACC), the vehicles in a convoy follow each other under adaptive cruise control (ACC) that is augmented by the sharing of preceding vehicle acceleration through the vehicle to vehicle communication in a feedforward control path. In general, the desired velocity optimization for vehicles in the convoy is based on fuel economy optimization, rather than driveability. This paper is a preliminary study on the impact of the desired velocity profile on the driveability characteristics of a convoy of vehicles and the controller gain impact on the driveability. A simple low-level longitudinal model of the vehicle has been used along with a PD type cruise controller and a generic spacing policy for ACC/CACC. The acceleration of the previous vehicle is available to the next vehicle as input, and the simulations are performed as Cooperative Adaptive Cruise Control of a convoy of vehicles. Individual vehicle acceleration profiles have been analyzed for driveability for two different velocity profiles that are followed in a stretch of 720 m between stop signs. The controller gains have been re-tuned based on the parameter space robust control PID approach for driveability and compared with the original gains. The US06 SFTP drive cycle has also been used for the comparison of the two different controller gain sets.

## Introduction

The development of autonomous vehicle technology is advancing every day, and its implementation in Personalized Rapid Transport (PRT) and Bus Rapid Transport (BRT) seem beneficial. There are several discussions happening in city planning departments on the idea of dedicated lanes for buses that operate autonomously. In this scenario, it is important to understand the driveability of a convoy of vehicles in the automated lane as passenger comfort is also an important issue. Driveability refers to the response of the vehicle to the driver input and the smoothness of motion. The intention is that passengers do not feel discomfort when traveling in the vehicle. Being an autonomous vehicle, the human driving is replaced by providing a lead vehicle velocity profile, which is calculated based on a velocity optimization algorithm. The optimization algorithm tries to find the optimized velocity profile based on maximum fuel economy. The optimization is usually not done for optimal driveability, and therefore it is important to study the impact of different velocity profiles on driveability considering that driverless cars are not necessarily passengerless.

Driveability is evaluated by experienced test drivers who observe the vehicle's dynamic response by providing various inputs such as throttle pedal tip-in/tip out, gear shifting and engine idling etc. During these tests, various behaviors such as jerk, vehicle shuffle, hunting, and oscillations are evaluated. The references to driveability measures by Wei *et al* [1] and Yard [2] provide a good insight about measuring different driveability evaluation metrics. There are also commercially available software packages like AVL Drive [3,4,5] that can be used to evaluate the acceleration profile of a vehicle using neural network algorithms, analyze the data from different vehicles and the ratings provided by different experienced drivers. This paper analyzes information from the acceleration signal and computes various parameters as mentioned by Wei *et al* [1], to arrive at the driveability measures. Signals below 10 Hz are not perceivable by humans as is mentioned by Schoeggl *et al* [5].

The convoy model used in this paper uses Cooperative Adaptive Cruise Control (CACC). In CACC, the vehicles receive information about the acceleration of the preceding vehicle and utilize this information to maintain the time headway between each vehicle. So, the acceleration behavior of the lead vehicle will affect the rest of the vehicles in the convoy. The paper sections are organized as follows. The next section on model development explains the equations used in vehicle dynamics modeling and longitudinal convoy modeling. The section on driveability presents different measures that are used for analyzing the vehicle acceleration signal and their calculation. The section on the desired velocity profile of the autonomous vehicles explains the reasoning behind choosing these different velocity profiles. The section on simulation results discusses the outcome of the convoy model simulated for different velocity profiles. Driveability metrics are presented to evaluate performance. The paper ends with conclusions and recommendations for future work.

## Model development

The models used here have been developed based on the CACC structure described in Gerrit et al [6]. A string of homogenous vehicles is being considered for the simulation as shown in Figure 1, given that the scenario is a dedicated bus lane, wherein one bus follows the other. The same analysis can be extended to a mixed traffic model also. Acceleration of each vehicle is represented as $a_i$, the standstill distance between two vehicles is given as $r_i$, and the length of each vehicle is given as $L_i$.

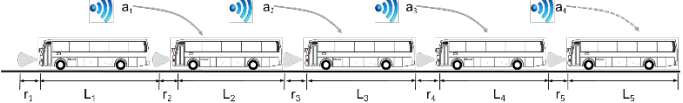

Figure 1. Architecture of the Five Vehicle Convoy

The main control objective of the Cooperative Adaptive Cruise Control model is to follow the preceding vehicle with desired time headway with the information of the acceleration of the preceding vehicle available as input. The vehicle model is a low-level closed-loop longitudinal vehicle dynamics, which can be represented for the $i^{th}$ vehicle as

$$G_i(s) = \frac{1}{s^2(\tau_i s + 1)} e^{-\phi_i s} \quad \text{for } i \geq 1 \quad (1)$$

The input $u_i(t)$ for the model $G_i(s)$ is desired acceleration. $1/\tau_i$ is the low-level closed loop bandwidth and $\phi_i$ is the delay in the actuator response and the internal communication delay that are lumped together in equation (1). In Figure 2, the acceleration signal from the previous vehicle $\ddot{x}_{i-1}$ is fed via a feedforward filter $F_i$. A standard PD type controller is used for tracking the desired acceleration, whose gain is given as $K_i(s)$. The PD controller is expressed as

$$K_i(s) = k_{P,i} + s k_{D,i} \quad \text{for } i \geq 1 \quad (2)$$

where $k_{D,i}$ and $k_{P,i}$ are the gains of the derivative and proportional controller for the $i^{th}$ vehicle.

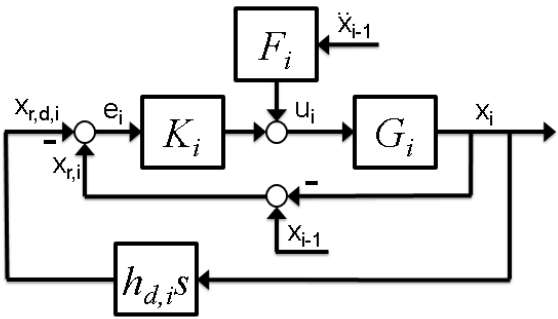

Figure 2: CACC Control block diagram

The spacing policy is used to calculate the desired relative position and consists of the constant distance and a velocity dependent part, and is given as,

$$x_{r,i}(t) = x_{r0,i} + h_{d,i} \dot{x}_i(t) \quad (3)$$

where $x_{r0,i} = L_i + r_i$, are the constant distance and the desired time headway is given as $h_{d,i}$ in seconds. The vehicle model parameters used are given in Table 1.

Table 1: Parameters used in the Model

| Parameters | L (m) | R (m) | $\phi_i$ (s) | $\tau$ (s) |
|---|---|---|---|---|
| Values | 10 | 10 | 0.1 | 0.5 |

## Driveability measures

Driveability is the perception of the driver about the quality of the drive and how smooth it is. In engineering terms, this can be broken down into some key measures, such as jerk, response time, initial bump, kick, vibration dose value (VDV) and stumble, as shown by Wei et al [1], Yard [2] and Jauch et al [9]. These measures are calculated by analysis of the acceleration signal to capture its characteristics at its different profile segments.

### Initial Response time
Initial response time is defined as the time it takes for the vehicle acceleration to reach 0.5 m/s². For a convoy, this initial response time means how quickly all the vehicles in the convoy start to move and reach the 0.5 m/s² acceleration value. The shorter the response time, the faster is the convoy response. It should be noted that each vehicle will have its own response time (depending on the powertrain configurations), and in this paper, since we are considering only homogenous vehicles, the response time of the individual vehicles is the same and the total convoy response time is governed by the lead vehicle velocity profile.

### Initial Bump
This is gradient of the acceleration rise, meaning the angle between the time axis and the initial acceleration profile slope. A higher initial slope means that the vehicle can accelerate quickly and this has a direct relationship to the response time. If the vehicle initial bump is higher, then the convoy will also accelerate faster and therefore the response time will be shorter.

### Kick
The kick is defined as the maximum drop in the acceleration after the vehicle attains its maximum acceleration. The higher the kick value, the more the driver feels the sharp jerk and resulting discomfort. A too steep acceleration rise will also lead to a higher drop in the acceleration and therefore the kick will be higher.

### Stumble
Stumble is defined as the drop in the acceleration when the torque is requested from the vehicle. This will also capture the behavior when the acceleration shuffles between positive and negative values, and the driver will feel that the vehicle might stall and might experience severe oscillations. The more the stumble, the worse is the driveability.

## *Jerk*

Jerk is calculated as the derivative of the acceleration, and the faster the acceleration changes, the more is the jerk. As mentioned in Wei *et al* [3], a value of 2 m/s$^3$ is an acceptable jerk. Note that jerk is sometimes calculated differently for driveability evaluation purposes.

## *Vibration Dose Value (VDV)*

It has been shown in Wei *et al* [1], that the VDV value gives a holistic indication of the vibration experienced by the human driver, who is in contact with the vibrating surface. This was mainly used as a measure in the industry to see the vibration experienced by the workers working with vibrating surfaces or machines. The acceleration signal is filtered between 1 Hz to 32 Hz using a time window between $t_0$ and final time $t_f$, as given by Horste [8]. The VDV calculation is made according to

$$VDV = \left[ \int_{t_0}^{t_f} \tilde{a}^4(t) dt \right]^{1/4} \quad (4)$$

where $\tilde{a}$ is the filtered acceleration value.

## Desired velocity profile for Autonomous vehicle

In this paper, we have tried two different desired velocity profiles to analyze the driveability of each vehicle in the convoy. The total distance is 720 m between two stop signs as shown in Figure 3 and is covered in different durations of time depending on the speed profile. The maximum speed limit for the stretch considered is 35 mph which corresponds to 15.4 m/s.

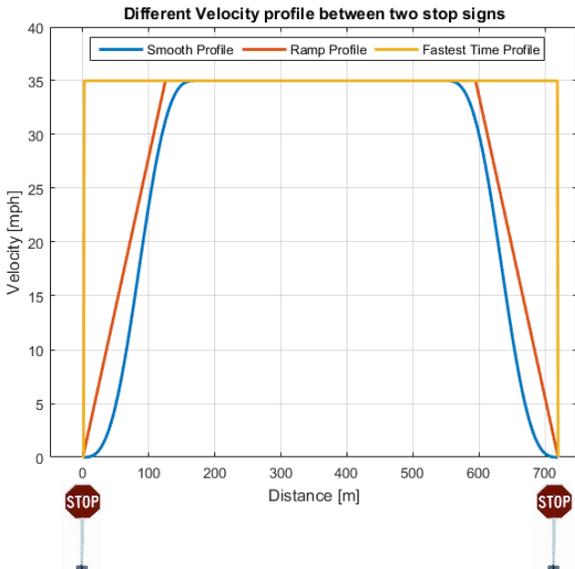

Figure 3. The two different desired velocity profiles between two stop signs

The ramp velocity profile increases the velocity to have a maximum acceleration of 3m/s$^2$. The second profile is the smooth velocity profile obtained by using cubic polynomials to achieve a smooth rise in the velocity. The step profile where the vehicle velocity reaches the speed limit instantly is given as a reference in this figure. This is not a practical scenario. We have used it to show the difference in the ramp velocity profile & smooth velocity in contrast to the theoretical fastest vehicle speed profile (shortest time). The durations of the velocity profiles have been tabulated in Table 2.

Table 2. Time taken for each velocity profile to cover 720m

| Velocity profiles | Time taken (s) |
|---|---|
| Step profile | 46.1582 |
| Ramp profile | 55.7001 |
| Smooth profile | 60.2573 |

Along with the theoretical profiles above, the simulation was also performed on the US06 drive cycle, which is the supplemental federal test procedure developed to represent the aggressive, high acceleration driving behavior of the driver.

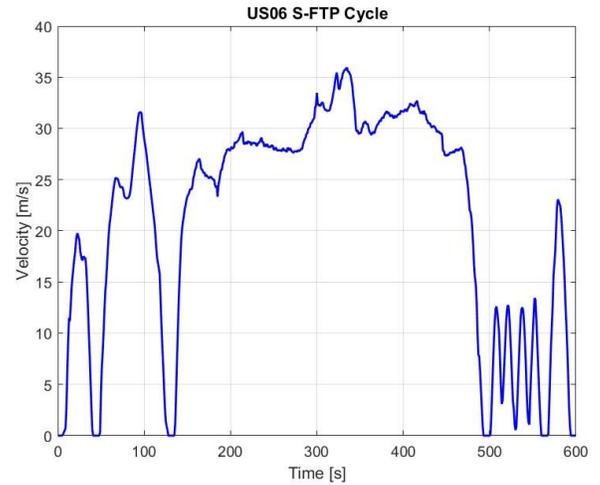

Figure 4. The US06 supplemental federal drive cycle contains high accelerations

As we see from Figure 4, the US06 drive cycle has very high acceleration peaks due to the steep rises and falls in velocity early in the cycle which thereby makes it a good candidate to use in the evaluation of CACC convoy driveability

## CACC controller parameters

Based on the literature review, the CACC controller gains were taken as $k_{D,i} = 2$ and $k_{P,i} = 4$ for the model, and we denote this as the original controller in this paper. The performance of the CACC controller depends heavily on the headway time gap and, therefore, the selection of the suitable headway time gap is important. According to Gerrit *et al* [6], the headway time gap of 0.6s has been achieved for best performance. In this paper, we have analyzed the model for 0.3s and 0.6s to find out the suitable headway time gap for driveability performance.

## Headway time gap simulation

In the vehicle convoy model, the time gap represents how closely the vehicles can follow each other. Initially, the headway time gap of 0.3s has been considered, in order to have the higher traffic density.

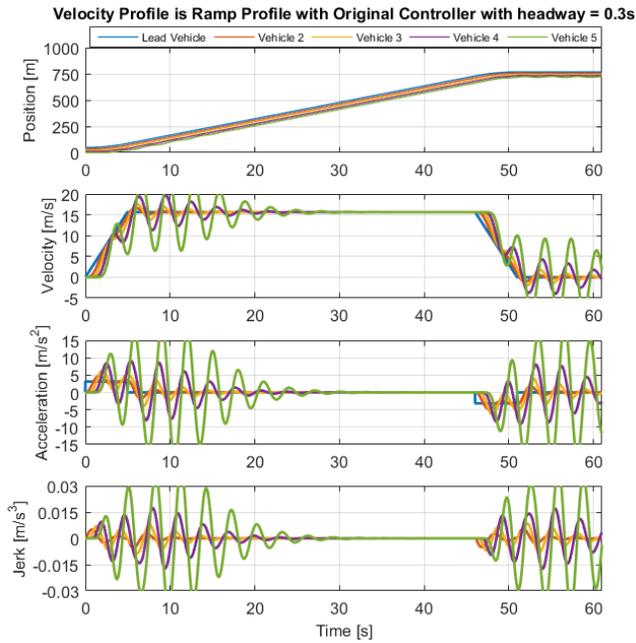

Figure 5. The convoy simulation for headway time gap of 0.3s for a desired ramp velocity profile.

In Figure 5, for the ramp velocity profile, the acceleration values are higher than 15 m/s$^2$ for one of the following vehicles and there are significant oscillations in the acceleration and velocity outputs of the convoy vehicles. These are undesirable behavior not suitable for convoy operation

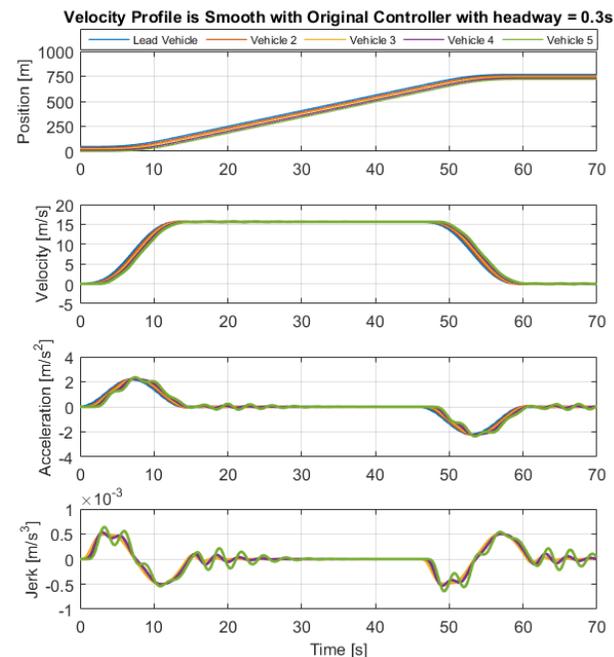

Figure 6. The convoy simulation for headway time gap of 0.3s for a desired smooth velocity profile.

. For the desired smooth velocity profile in Figure 6, although we find that the acceleration values are within the acceptable range of +2m/s$^2$ and -2.5m/s$^2$, it can be seen that the acceleration of the last vehicle starts to oscillate, which is undesirable in a convoy. This leads to the conclusion that headway time gap of 0.3s is not feasible for the type of vehicles considered. Hence, we chose a 0.6 s headway time gap and the corresponding driveability analysis is discussed in the next section.

## Evaluation of driveability performance of the CACC controller with original gains

In order to assess the driveability performance of the CACC controller, a simulation was performed with an input of the theoretical profile of a ramp and a smooth velocity profile and the resulting acceleration signal is analyzed for the driveability parameters as mentioned in the previous section.

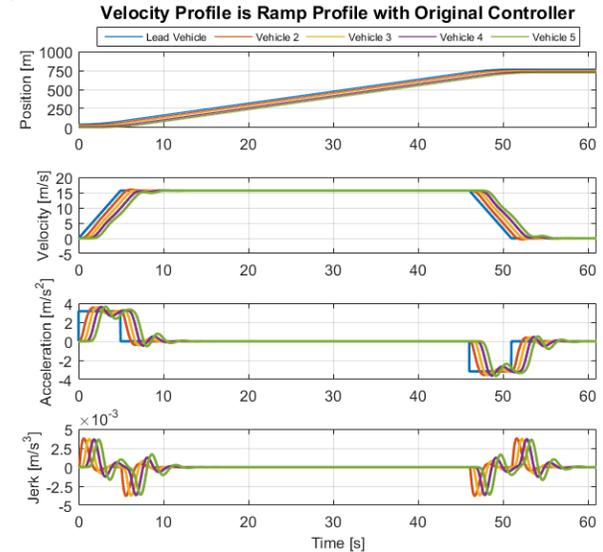

Figure 7. The convoy simulation for headway time gap of 0.6s for a desired ramp velocity profile with the original controller gains.

In Figure 7, we find that the acceleration values are slightly higher than the 3 m/s$^2$ and the jerk values are closer to $4 \times 10^{-3}$ m/s$^3$. The driveability measure of the convoy is shown in Figure 8.

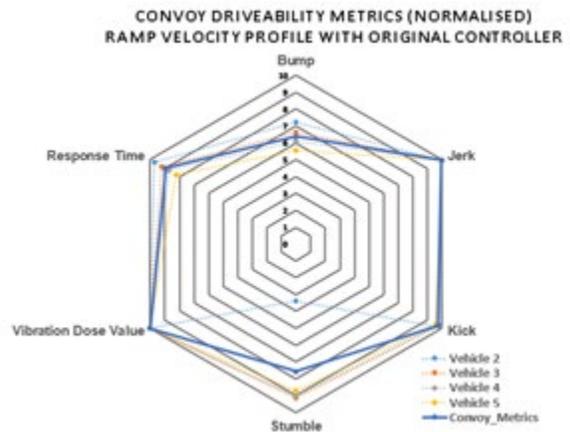

Figure 8. Spider plot of driveability analysis for headway time gap of 0.6s for a desired ramp velocity profile with the original controller gains.

This spider plot shows all driveability measures for each vehicle in the convoy on the same plot. This allows the reader to comprehend how the driveability metrics change as we go from one vehicle to the next in the convoy. The spider graph shows the normalized value of the driveability metrics which mean that 10 is the best value and 0 is the worst one. The limits for normalization are chosen from the literature.

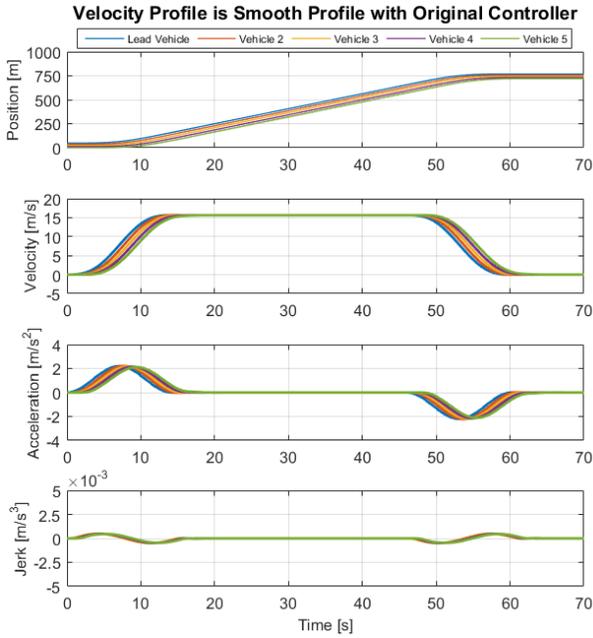

Figure 9. The convoy simulation for headway time gap of 0.6s for a desired smooth velocity profile with the original controller gains.

For the ramp velocity profile, the driveability measures show that the jerk, Vibration dose values, kick and responses time are satisfactory for the convoy. The bump response is getting poorer as the vehicle position in the convoy changes. This is mainly because of the delay between the start of the convoy and the time taken for the last vehicle to start. Similarly, the stumble is higher for the first vehicle, as the vehicle makes an instant start to follow the velocity profile thereby has a higher acceleration causing it to stumble and the following vehicles make a gradual start and thereby their stumble is lesser as the vehicle position increases.

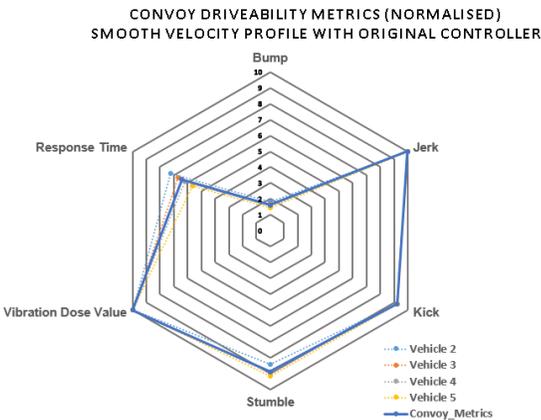

Figure 10. Spider plot of driveability analysis for headway time gap of 0.6s for a desired smooth velocity profile with the original controller gains.

In Figure 9, we find that the acceleration values for the smooth profile are relatively lower than that of the ramp profile, and they reach a maximum of 2 m/s² while the jerk is also substantially lower, and has a peak of about 0.95x10⁻³ m/s³. The corresponding driveability metric spider plot in Figure 10 has an improved kick and stumble when compared to the ramp profile. The response time and the bump values got lower since the smooth profile meant a delayed and slow increase in the acceleration, thereby affecting the response time and bump.

## D-Stability design for controller gains

The parameter space approach based PD controller design is applied.to robust Cooperative Adaptive Cruise Controller. The theoretical background about the parameter space approach and the corresponding generic equations can be found in references [7-9,12]. The controller parameter space is obtained considering D-stability requirements shown in Figure 11 for the two free coefficients of a PD controller as the proportional gain $k_{P,i}$ and the derivative gain $k_{D,i}$. Figure 11 can be interpreted as constraining the poles of the closed loop PD-controlled system to lie in the D-Stable region of the desired settling time ($\partial_1$), a desired minimum damping ratio ($\partial_2$) and the desired maximum bandwidth ($\partial_3$).

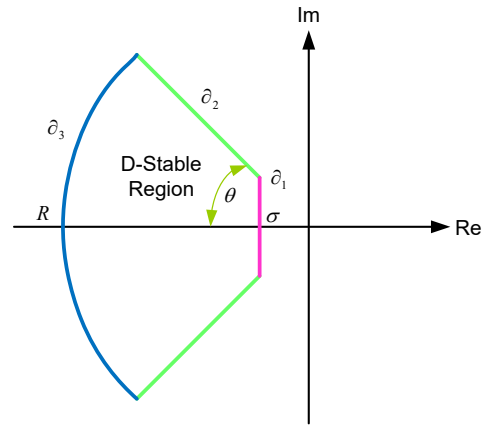

Figure 11: The D-stable region in Parameter space approach

Robust PD coefficients satisfying D-stability are chosen from the overall calculated parameter space region. The design criteria are considered by increasing the damping ratio to suppress the acceleration oscillations which is beneficial in damping the response further to improve the driveability characteristics. The design parameters in Figure 11 are, settling time constraint of $\sigma = 0.5$, bandwidth constraint of 20, and damping ratio constraint of at least 0.866, which corresponds to θ=30°. Figure 12 shows the PD controller parameter space that satisfies the design criteria and the controller gains are chosen from this region as $k_{D,i}$=0.6575 and $k_{P,i}$=0.1609 as shown by the small circle.

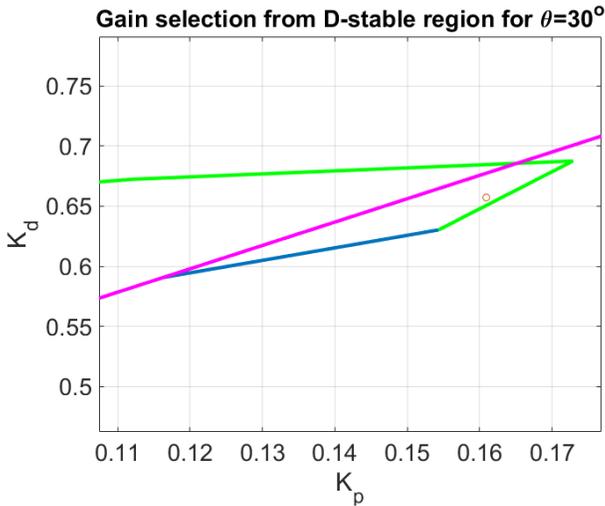

Figure 12: CACC Controller gain selection from the region satisfying the D-stable conditions for design 1 with damping ratio of 0.866, θ=30°

The pole locations for the corresponding gain values are shown in Figure 13. The resulting complex conjugate poles are seen to be located very close to the damping ratio constraint boundary. There are also two real poles. One of these is close to the settling time constraint boundary while the other one is close to the bandwidth constraint boundary. All poles are within the D-stability region.

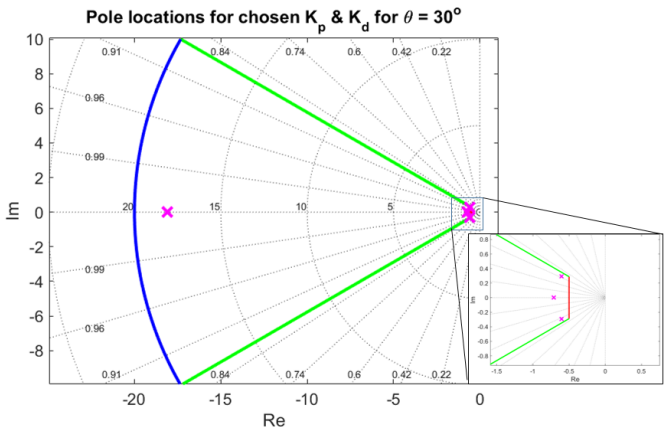

Figure 13: Pole locations for the selected CACC Controller gain selection from the region satisfying the D-stable conditions for design 2 with θ=30°

## Simulation Results of new controller

The driveability analysis has been performed by simulating the vehicle convoy model in the CACC mode, with the new controller gains obtained from the D-stability analysis with the headway time gap of 0.6 s. The simulation is performed with the same ramp and smooth velocity profiles as earlier, to be able to make a comparison with the original controller. The spider plot condenses the different normalized driveability parameters for each simulation and gives a consolidated understanding [13]

*Ramp Profile Results*
The lead vehicle acceleration increases due to the change in the velocity, stays constant and later decreases in the ramp velocity profile simulation shown in Figure 14.

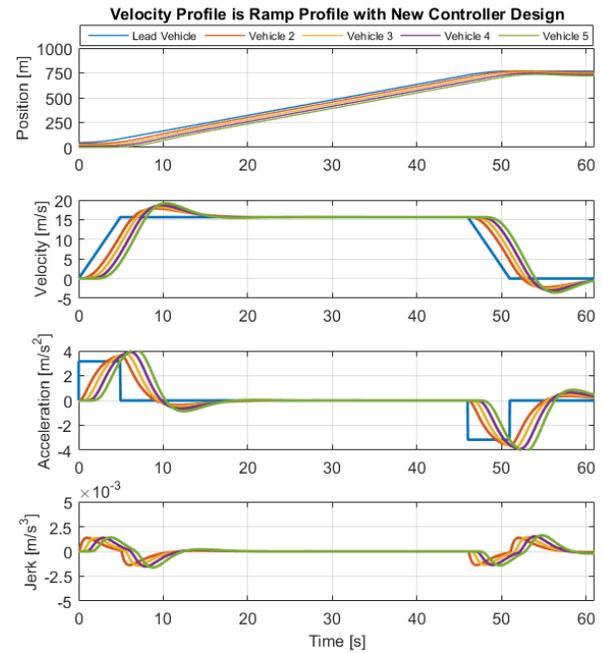

Figure 14. The position, velocity, acceleration and jerk of the vehicles in the convoy for a Ramp velocity profile

The acceleration is in the range of ±4 m/s$^2$ and there is an overshoot followed by an undershoot in the acceleration signals. The profile is still aggressive given that the comfortable acceleration range is between ±2.5 m/s$^2$. The corresponding spider plot for the driveability metrics is shown in a normalized manner in Figure 15. It can be seen that the response time and bump have improved compared to the ramp profile for the original controller. The maximum jerk value experienced in this profile for the new controller gains is 1.685x10$^{-3}$ m/s$^3$ which is significantly less than that obtained with the original controller gains.

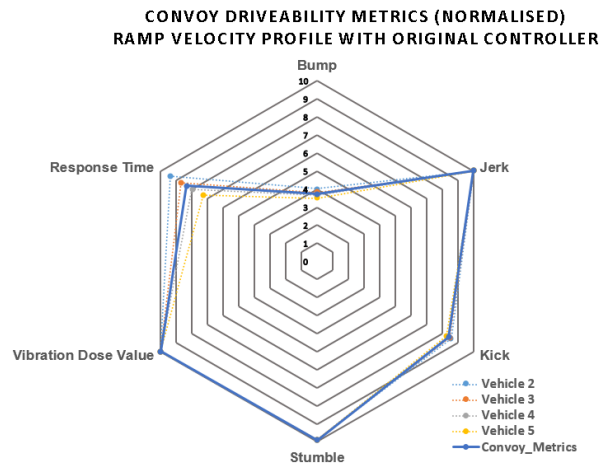

Figure 15. Spider plot of driveability analysis for headway time gap of 0.6s for a desired smooth velocity profile with new controller gains

*Smooth Polynomial Profile Results*
Figure 16, shows the response to the smooth polynomial desired velocity profile. It can be noted that the acceleration profile increases smoothly to a value of +2.3 m/s$^2$ and reduces to a value of -0.5 m/s$^2$ during the stopping phase. This is a very comfortable compared to the ramp profile and also compared to what was achieved with the original controller gains. The jerk values are significantly lower than the ramp velocity profile and what was achieved with the original controller gains. The maximum value of the jerk is $0.5 \times 10^{-3}$ m/s$^3$. This indicates that each vehicle in the convoy feels a smooth acceleration and deceleration without experiencing an appreciable jerk.

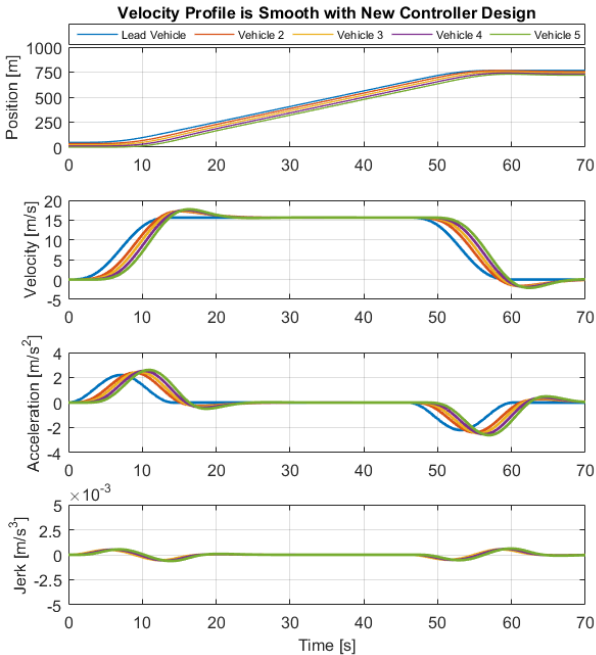

Figure 16. The position, velocity, acceleration and jerk of the vehicles in the convoy for a smooth polynomial desired velocity profile

Since the controller gains were designed to be less aggressive, we are seeing fewer oscillations. This is seen clearly in the driveability evaluation spider plot of Figure 17. The kick, jerk, stumble and VDV values are lower as compared to the ramp profile.

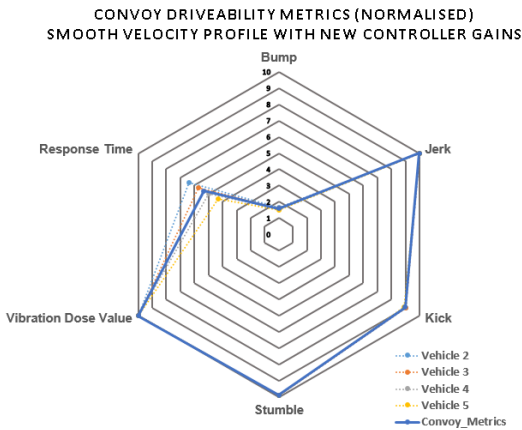

Figure 17. Spider plot of driveability measures for all the vehicles in the convoy for a smooth polynomial velocity profile with new controller gains.

It should be noted that the response time of the convoy is low and is predominantly due to the acceleration that rises slowly and gradually. This is also seen in the normalized bump value which is also lower when compared to that of the other two profiles.

### US06 S-FTP Results

The US06 drive cycle was simulated and analyzed for the convoy driveability with both the original controller gains and the new controller gains.

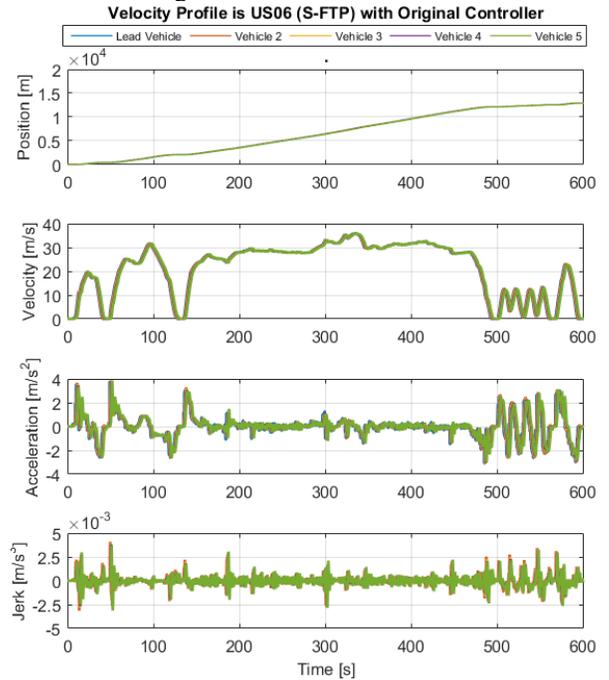

Figure 18. Driveability analysis on the US06 drive cycle

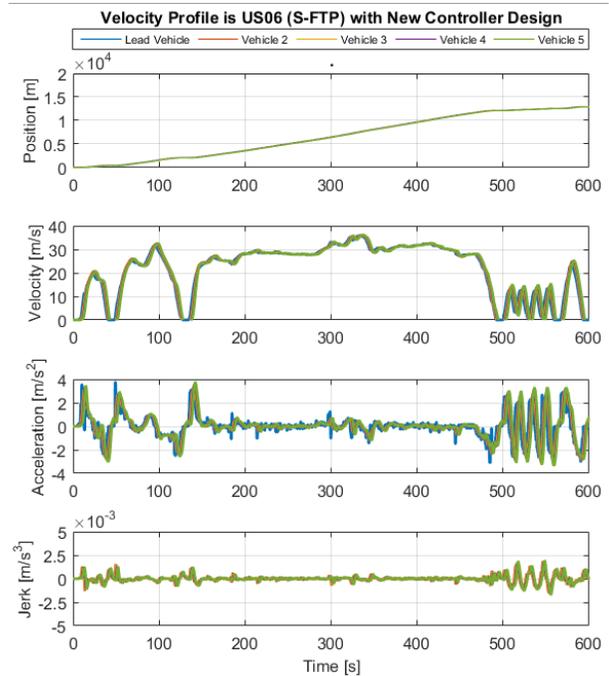

Figure 19. Driveability analysis on the US06 drivecycle with the new controller gains for the whole cycle

The results are shown in Figure 18 for the original controller gains and in Figure 19 for the new controller gains based on parameter space D-stability tuning. A comparison of Figures 18 and 19 shows that jerk is reduced significantly when the new controller gains are used. In order to better understand the impact on driveability during the high acceleration region, the first 60 s of the simulated with the original controller gains and new controller gains will be studied next. The corresponding plots are shown in Figure 20 for the original controller and Figure 21 for the new controller.

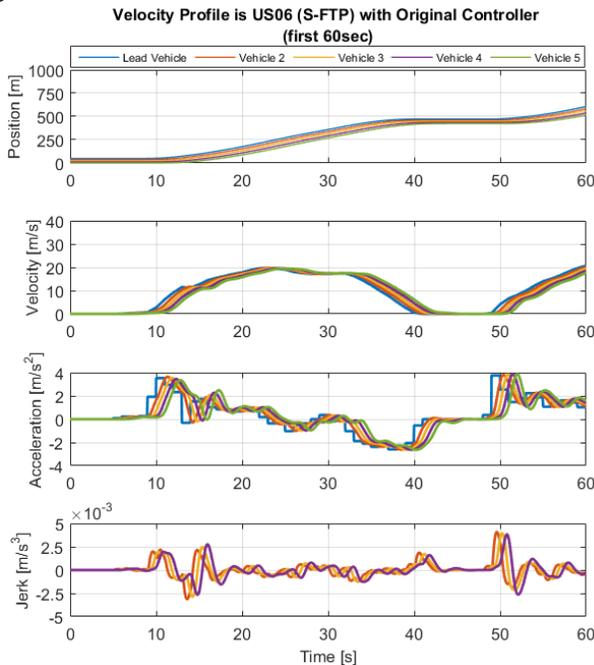

Figure 20. Driveability analysis on the US06 drivecycle focused on the first two high peak acceleration (0-60sec) with original controller gains

For this 60 s high acceleration comparison, we can use the peak jerk value as the indication for the driveability. It can be seen that during the first 60 s of the drive cycle, the peak jerk value for the original controller is at $2.5 \times 10^{-3}$ m/s$^3$ compared to $1.01 \times 10^{-3}$ m/s$^3$ for the new controller gains. This indicates that new controller gain is effective in improving the driveability performance of the convoy.

## Conclusions

The results reported in the paper show that the controller gains tuned for performance and the road profiles have a significant impact on the driveability. Smoother profiling of the lead vehicle's desired velocity and the tuning of the controller gain to be less aggressive will be beneficial for convoy driveability in a cooperative adaptive cruise control model. This also shows that a smooth desired velocity profile like the polynomial profile is necessary to improve driveability although it will result in a larger travel time. The modeling and analysis presented in this paper have been setup in such a way, that we can extend our results to different controller design methods and to heterogeneous convoys which will be a part of the future work. Future work can also focus on integrating or using the drivability and parameter space control design approaches in this paper with other control, AV, CV, ADAS, automotive control and other topics like those in references [14-75].

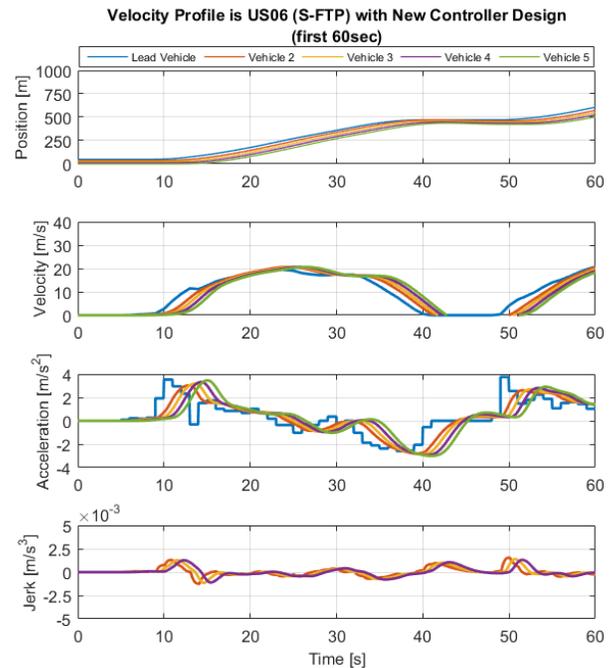

Figure 21. Driveability analysis on the US06 drivecycle, focused on the first two high peak acceleration (0-60sec) with new controller gains

## References

1. Wei, X., Rizzoni, G. et al. Objective Metrics of Fuel Economy, Performance and Driveability - A Review. SAE Technical Paper, 2004-01-1338, 2004, doi: 10.4271/2004-01-1338.

2. Yard, M. Control and Drive Quality Refinement of a Parallel-Series Plug-in Hybrid Electric Vehicle. Master thesis, Ohio State University, 2014.

3. List, H. O. and Schoeggl, P. Objective Evaluation of Vehicle Driveability. SAE Technical Paper, no. 980204, 1998.

4. Schoeggl, P. and Ramschak, E. Vehicle Driveability Assessment using Neural Networks for Development, Calibration and Quality Tests. SAE Technical Paper, no.2000-01-0702, 2000.

5. Schoeggl, P., Kriegler, W. and Bogner, E. Virtual Optimization of Vehicle and Powertrain Parameters with Consideration of Human Factors. SAE Technical Paper, no. 2005-01-1945, 2005.

6. Gerrit J. L. Naus, René P. A. Vugts, Jeroen Ploeg, Marinus (René) J. G. van de Molengraft, and Maarten Steinbuch, Senior Member, IEEE, String-Stable CACC Design and Experimental Validation: A Frequency-Domain Approach, IEEE Transactions on Vehicular Technology, vol. 59, no. 9, November 2010

7. Ackermann, J., Blue, P., Bunte, T., Güvenç, L., Kaesbauer, D., Kordt, M., Muhler, M., Odenthal, D., 2002, "Robust Control: The Parameter Space Approach, London", U.K.: Springer-Verlag.

8. Güvenç, L. and Ackermann, J., 2001, "Links between the parameter space and frequency domain methods of robust control", Int. J. of Robust and Nonlinear Control, vol. 11, pp. 1435-1453.

9. Demirel, B. and Güvenç, L., 2010, "Parameter space design of repetitive controllers for satisfying a robust performance requirement", IEEE Trans. on Automatic Control, vol. 55, no. 8, pp. 1893-1899.

10. C. Jauch, K. Bovee, S. Tamilarasan, L. Guvenc, and G. Rizzoni, "Modeling of the OSU EcoCAR 2 vehicle for Driveability Analysis," in E-COSM'15 IFAC Workshop on Engine and Powertrain Control, Simulation and Modeling, 2015.

11. Horste, K. Objective Measurement of Automatic Transmission Shift Feel Using Vibration Dose Value. SAE Technical Paper, No. 951373, 1995.

12. Ackermann, J., 1980, "Parameter Space Design of Robust Control Systems," IEEE Transactions on Automatic Control, Vol. 25, pp. 1058-1072.

13. Jauch, C., Tamilarasan, S., Bovee, K., Guvenc, L., Rizzoni, G. Design and verification of drivability improving control for the EcoCAR 2 hybrid electric vehicle, Proceedings of the American Control Conference, 2016

14. Emirler, M.T., Uygan, I.M.C., Aksun Guvenc, B., Guvenc, L., 2014, "Robust PID Steering Control in Parameter Space for Highly Automated Driving," International Journal of E. Kural, Vol. 2014, Article ID 259465.

15. Emirler, M.T., Wang, H., Aksun Guvenc, B., Guvenc, L., 2015, "Automated Robust Path Following Control based on Calculation of Lateral Deviation and Yaw Angle Error," ASME Dynamic Systems and Control Conference, DSC 2015, October 28-30, Columbus, Ohio, U.S.

16. Wang, H., Tota, A., Aksun-Guvenc, B., Guvenc, L., 2018, "Real Time Implementation of Socially Acceptable Collision Avoidance of a Low Speed Autonomous Shuttle Using the Elastic Band Method," IFAC Mechatronics Journal, Volume 50, April 2018, pp. 341-355.

17. Zhou, H., Jia, F., Jing, H., Liu, Z., Guvenc, L., 2018, "Coordinated Longitudinal and Lateral Motion Control for Four Wheel Independent Motor-Drive Electric Vehicle," IEEE Transactions on Vehicular Technology, Vol. 67, No 5, pp. 3782-379.

18. Emirler, M.T., Kahraman, K., Senturk, M., Aksun Guvenc, B., Guvenc, L., Efendioglu, B., 2015, "Two Different Approaches for Lateral Stability of Fully Electric Vehicles," International Journal of Automotive Technology, Vol. 16, Issue 2, pp. 317-328.

19. S Zhu, B Aksun-Guvenc, Trajectory Planning of Autonomous Vehicles Based on Parameterized Control Optimization in Dynamic on-Road Environments, Journal of Intelligent & Robotic Systems, 1-13, 2020.

20. Gelbal, S.Y., Cantas, M.R, Tamilarasan, S., Guvenc, L., Aksun-Guvenc, B., 2017, "A Connected and Autonomous Vehicle Hardware-in-the-Loop Simulator for Developing Automated Driving Algorithms," IEEE Systems, Man and Cybernetics Conference, Banff, Canada.

21. Boyali A., Guvenc, L., 2010, "Real-Time Controller Design for a Parallel Hybrid Electric Vehicle Using Neuro-Dynamic Programming Method," IEEE Systems, Man and Cybernetics, İstanbul, October 10-13, pp. 4318-4324.

22. Kavas-Torris, O., Cantas, M.R., Gelbal, S.Y., Aksun-Guvenc, B., Guvenc, L., "Fuel Economy Benefit Analysis of Pass-at-Green (PaG) V2I Application on Urban Routes with STOP Signs," Special Issue on Safety and Standards for CAV, International Journal of Vehicle Design, in press.

23. Yang, Y., Ma, F., Wang, J., Zhu, S., Gelbal, S.Y., Kavas-Torris, O., Aksun-Guvenc, B., Guvenc, L., 2020, "Cooperative Ecological Cruising Using Hierarchical Control Strategy with Optimal Sustainable Performance for Connected Automated Vehicles on Varying Road Conditions," Journal of Cleaner Production, Vol. 275, in press.

24. Hartavi, A.E., Uygan, I.M.C., Guvenc, L., 2016, "A Hybrid Electric Vehicle Hardware-in-the-Loop Simulator as a Development Platform for Energy Management Algorithms," International Journal of Vehicle Design, Vol. 71, No. 1/2/3/4, pp. 410-420.

25. H. Gunther, O. Trauer, and L. Wolf, "The potential of collective perception in vehicular ad-hoc networks," 14th International Conference on ITS Telecommunications (ITST), 2015, pp. 1–5.

26. R. Miucic, A. Sheikh, Z. Medenica, and R. Kunde, "V2x applications using collaborative perception," IEEE 88th Vehicular Technology Conference (VTC-Fall), 2018, pp. 1–6.

27. B. Mourllion, A. Lambert, D. Gruyer, and D. Aubert, "Collaborative perception for collision avoidance," IEEE International Conference on Networking, Sensing and Control, 2004, vol. 2, pp. 880–885.

28. Gelbal, S.Y., Aksun-Guvenc, B., Guvenc, L., 2020, "Elastic Band Collision Avoidance of Low Speed Autonomous Shuttles with Pedestrians," International Journal of Automotive Technology, Vol. 21, No. 4, pp. 903-917.

29. E. Kural and B. Aksun-Guvenc, "Model Predictive Adaptive Cruise Control," in *IEEE International Conference on Systems, Man, and Cybernetics*, Istanbul, Turkey, October 10-13, 2010.

30. L. Guvenc, B. Aksun- Guvenc, B. Demirel and M. Emirler, Control of Mechatronic Systems, London: the IET, ISBN: 978-1-78561-144-5, 2017.

31. B. Aksun-Guvenc, L. Guvenc, E. Ozturk and T. Yigit, "Model Regulator Based Individual Wheel Braking Control," in *IEEE Conference on Control Applications*, Istanbul, June 23-25, 2003.

32. B. Aksun-Guvenc and L. Guvenc, "The Limited Integrator Model Regulator and its Use in Vehicle Steering Control," *Turkish Journal of Engineering and Environmental Sciences,* pp. pp. 473-482, 2002.

33. S. K. S. Oncu, L. Guvenc, S. Ersolmaz, E. Ozturk, A. Cetin and M. Sinal, "Steer-by-Wire Control of a Light Commercial Vehicle Using a Hardware-in-the-Loop Setup," in *IEEE Intelligent Vehicles Symposium*, June 13-15, pp. 852-859, 2007.

34. M. Emirler, K. Kahraman, M. Senturk, B. Aksun-Guvenc, L. Guvenc and B. Efendioglu, "Two Different Approaches for Lateral Stability of Fully Electric Vehicles," *International*

*Journal of Automotive Technology,* vol. 16, no. 2, pp. 317-328, 2015.

35. S. Y. Gelbal, M. R. Cantas, B. Aksun-Guvenc, L. Guvenc, G. Surnilla, H. Zhang, M. Shulman, A. Katriniok and J. Parikh, "Hardware-in-the-Loop and Road Testing of RLVW and GLOSA Connected Vehicle Applications," in *SAE World Congress Experience*, 2020.

36. Gelbal, S.Y., Cantas, M.R, Tamilarasan, S., Guvenc, L., Aksun-Guvenc, B., "A Connected and Autonomous Vehicle Hardware-in-the-Loop Simulator for Developing Automated Driving Algorithms," in *IEEE Systems, Man and Cybernetics Conference*, Banff, Canada, 2017.

37. S. Y. Gelbal, B. Aksun-Guvenc and L. Guvenc, "SmartShuttle: A Unified, Scalable and Replicable Approach to Connected and Automated Driving in a SmartCity," in *Second International Workshop on Science of Smart City Operations and Platforms Engineering (SCOPE)*, Pittsburg, PA, 2017.

38. B. Demirel and L. Guvenc, "Parameter Space Design of Repetitive Controllers for Satisfying a Mixed Sensitivity Performance Requirement," *IEEE Transactions on Automatic Control,* vol. 55, pp. 1893-1899, 2010.

39. B. Aksun-Guvenc and L. Guvenc, "Robust Steer-by-wire Control based on the Model Regulator," in *IEEE Conference on Control Applications*, 2002.

40. B. Orun, S. Necipoglu, C. Basdogan and L. Guvenc, "State Feedback Control for Adjusting the Dynamic Behavior of a Piezo-actuated Bimorph AFM Probe," *Review of Scientific Instruments,* vol. 80, no. 6, 2009.

41. L. Guvenc and K. Srinivasan, "Friction Compensation and Evaluation for a Force Control Application," *Journal of Mechanical Systems and Signal Processing,* vol. 8, no. 6, pp. 623-638.

42. Guvenc, L., Srinivasan, K., 1995, "Force Controller Design and Evaluation for Robot Assisted Die and Mold Polishing," Journal of Mechanical Systems and Signal Processing, Vol. 9, No. 1, pp. 31-49.

43. M. Emekli and B. Aksun-Guvenc, "Explicit MIMO Model Predictive Boost Pressure Control of a Two-Stage Turbocharged Diesel Engine," IEEE Transactions on Control Systems Technology, vol. 25, no. 2, pp. 521-534, 2016.

44. Aksun-Guvenc, B., Guvenc, L., 2001, "Robustness of Disturbance Observers in the Presence of Structured Real Parametric Uncertainty," Proceedings of the 2001 American Control Conference, June, Arlington, pp. 4222-4227.

45. Guvenc, L., Ackermann, J., 2001, "Links Between the Parameter Space and Frequency Domain Methods of Robust Control," International Journal of Robust and Nonlinear Control, Special Issue on Robustness Analysis and Design for Systems with Real Parametric Uncertainties, Vol. 11, no. 15, pp. 1435-1453.

46. Demirel, B., Guvenc, L., 2010, "Parameter Space Design of Repetitive Controllers for Satisfying a Mixed Sensitivity Performance Requirement," IEEE Transactions on Automatic Control, Vol. 55, No. 8, pp. 1893-1899.

47. Ding, Y., Zhuang, W., Wang, L., Liu, J., Guvenc, L., Li, Z., 2020, "Safe and Optimal Lane Change Path Planning for Automated Driving," IMECHE Part D Passenger Vehicles, Vol. 235, No. 4, pp. 1070-1083, doi.org/10.1177/0954407020913735.

48. T. Hacibekir, S. Karaman, E. Kural, E. S. Ozturk, M. Demirci and B. Aksun Guvenc, "Adaptive headlight system design using hardware-in-the-loop simulation,"2006 IEEE Conference on Computer Aided Control System Design, 2006 IEEE International Conference on Control Applications, 2006 IEEE International Symposium on Intelligent Control, Munich, Germany, 2006, pp. 915-920, doi: 10.1109/CACSD-CCA-ISIC.2006.4776767.

49. Guvenc, L., Aksun-Guvenc, B., Emirler, M.T. (2016) "Connected and Autonomous Vehicles," Chapter 35 in Internet of Things/Cyber-Physical Systems/Data Analytics Handbook, Editor: H. Geng, Wiley.

50. Emirler, M.T.; Guvenc, L.; Guvenc, B.A. Design and Evaluation of Robust Cooperative Adaptive Cruise Control Systems in Parameter Space. International Journal of Automotive Technology 2018, 19, 359–367, doi:10.1007/s12239-018-0034-z.

51. Gelbal, S.Y.; Aksun-Guvenc, B.; Guvenc, L. Collision Avoidance of Low Speed Autonomous Shuttles with Pedestrians. International Journal of Automotive Technology 2020, 21, 903–917, doi:10.1007/s12239-020-0087-7.

52. Zhu, S.; Gelbal, S.Y.; Aksun-Guvenc, B.; Guvenc, L. Parameter-Space Based Robust Gain-Scheduling Design of Automated Vehicle Lateral Control. IEEE Transactions on Vehicular Technology 2019, 68, 9660–9671, doi:10.1109/TVT.2019.2937562.

53. Yang, Y.; Ma, F.; Wang, J.; Zhu, S.; Gelbal, S.Y.; Kavas-Torris, O.; Aksun-Guvenc, B.; Guvenc, L. Cooperative Ecological Cruising Using Hierarchical Control Strategy with Optimal Sustainable Performance for Connected Automated Vehicles on Varying Road Conditions. Journal of Cleaner Production 2020, 275, 123056, doi:10.1016/j.jclepro.2020.123056.

54. Cebi, A., Guvenc, L., Demirci, M., Kaplan Karadeniz, C., Kanar, K., Guraslan, E., 2005, "A Low Cost, Portable Engine ECU Hardware-In-The-Loop Test System," Mini Track of Automotive Control, IEEE International Symposium on Industrial Electronics Conference, Dubrovnik, June 20-23.

55. Ozcan, D., Sonmez, U., Guvenc, L., 2013, "Optimisation of the Nonlinear Suspension Characteristics of a Light Commercial Vehicle," International Journal of Vehicular Technology, Vol. 2013, ArticleID 562424.

56. Ma, F., Wang, J., Yu, Y., Zhu, S., Gelbal, S.Y., Aksun-Guvenc, B., Guvenc, L., 2020, "Stability Design for the Homogeneous Platoon with Communication Time Delay," Automotive Innovation, Vol. 3, Issue 2, pp. 101-110.

57. Tamilarasan, S., Jung, D., Guvenc, L., 2018, "Drive Scenario Generation Based on Metrics for Evaluating an Autonomous Vehicle Speed Controller For Fuel Efficiency Improvement," WCX18: SAE World Congress Experience, April 10-12, Detroit, Michigan, Autonomous Systems, SAE Paper Number 2018-01-0034.

58. Tamilarasan, S., Guvenc, L., 2017, "Impact of Different Desired Velocity Profiles and Controller Gains on Convoy Drivability of Cooperative Adaptive Cruise Control Operated Platoons," WCX17: SAE World Congress Experience, April 4-6, Detroit, Michigan, SAE paper number 2017-01-0111.

59. Altay, I., Aksun Guvenc, B., Guvenc, L., 2013, "Lidar Data Analysis for Time to Headway Determination in the DriveSafe Project Field Tests," International Journal of Vehicular Technology, Vol. 2013, ArticleID 749896.

60. Coskun, F., Tuncer, O., Karsligil, E., Guvenc, L., 2010, "Vision System Based Lane Keeping Assistance Evaluated in a Hardware-in-the-Loop Simulator," ASME ESDA Engineering Systems Design and Analysis Conference, İstanbul, July 12-14.

61. Aksun Guvenc, B., Guvenc, L., Yigit, T., Ozturk, E.S., 2004, "Coordination Strategies For Combined Steering and Individual Wheel Braking Actuated Vehicle Yaw Stability Control," 1st IFAC Symposium on Advances in Automotive Control, Salerno, April 19-23.

62. Emirler, M.T., Kahraman, K., Senturk, M., Aksun Guvenc, B., Guvenc, L., Efendioglu, B., 2013, "Estimation of Vehicle Yaw Rate Using a Virtual Sensor," International Journal of Vehicular Technology, Vol. 2013, ArticleID 582691.

63. Bowen, W., Gelbal, S.Y., Aksun-Guvenc, B., Guvenc, L., 2018, "Localization and Perception for Control and Decision Making of a Low Speed Autonomous Shuttle in a Campus Pilot Deployment," SAE International Journal of Connected and Automated Vehicles, doi: 10.4271/12-01-02-0003, Vol. 1, Issue 2, pp. 53-66.

64. Oncu Ararat, Bilin Aksun Guvenc, "Development of a Collision Avoidance Algorithm Using Elastic Band Theory," IFAC Proceedings Volumes, Volume 41, Issue 2, 2008, Pages 8520-8525.

65. Mumin Tolga Emirler, Haoan Wang, Bilin Aksun Guvenc, "Socially Acceptable Collision Avoidance System for Vulnerable Road Users," IFAC-PapersOnLine, Volume 49, Issue 3, 2016, Pages 436-441.

66. Cao, X., Chen, H., Gelbal, S.Y., Aksun-Guvenc, B., Guvenc, L., 2023, "Vehicle-in-Virtual-Environment (VVE) Method for Autonomous Driving System Development, Evaluation and Demonstration," Sensors, Vol. 23, 5088, https://doi.org/10.3390/s23115088.

67. Kavas-Torris, O., Gelbal, S.Y., Cantas, M.R., Aksun-Guvenc, B., Guvenc, L., "V2X Communication Between Connected and Automated Vehicles (CAVs) and Unmanned Aerial Vehicles (UAVs)," Sensors, Vol. 22, 8941. https://doi.org/10.3390/s22228941.

68. Meneses-Cime, K., Aksun-Guvenc, B., Guvenc, L., 2022, "Optimization of On-Demand Shared Autonomous Vehicle Deployments Utilizing Reinforcement Learning," Sensors, Vol. 22, 8317. https://doi.org/10.3390/s22218317.

69. Guvenc, L., Aksun-Guvenc, B., Zhu, S., Gelbal, S.Y., 2021, Autonomous Road Vehicle Path Planning and Tracking Control, Wiley / IEEE Press, Book Series on Control Systems Theory and Application, New York, ISBN: 978-1-119-74794-9.

70. Bilin Aksun Guvenc, Levent Guvenc, Tevfik Yigit, Eyup Serdar Ozturk, "Coordination Strategies for Combined Steering and Individual Wheel Braking Actuated Vehicle Yaw Stability Control," IFAC Proceedings Volumes, Volume 37, Issue 22, 2004, Pages 85-90.

71. Ackermann, J., Blue, P., Bunte, T., Guvenc, L., Kaesbauer, D., Kordt, M., Muhler, M., Odenthal, D., 2002, Robust Control: the Parameter Space Approach, Springer Verlag, London, ISBN: 978-1-85233-514-4.

72. Gozu, M., Ozkan, B. & Emirler, M.T. Disturbance Observer Based Active Independent Front Steering Control for Improving Vehicle Yaw Stability and Tire Utilization. Int.J Automot. Technol. 23, 841–854 (2022). https://doi.org/10.1007/s12239-022-0075-1.

73. Tamilarasan, Santhosh, and Levent Guvenc. "Impact of Different Desired Velocity Profiles and Controller Gains on Convoy Driveability of Cooperative Adaptive Cruise Control Operated Platoons." SAE International Journal of Passenger Cars - Mechanical Systems, vol. 10, no. 1, Apr. 2017, pp. 93+.

74. V. Sankaranarayanan *et al.*, "Observer based semi-active suspension control applied to a light commercial vehicle," *2007 IEEE/ASME international conference on advanced intelligent mechatronics*, Zurich, Switzerland, 2007, pp. 1-7, doi: 10.1109/AIM.2007.4412539.

75. Gao, Y., Shen, Y., Yang, Y., Zhang, W., Guvenc, L., 2019, "Modeling, Verification and Analysis of Articulated Steer Vehicles and a New Way to Eliminate Jack-Knife and Snaking Behavior," International Journal of Heavy Vehicle Systems, Vol. 26, Issue: 3-4, pp. 375-404.